\documentclass[aps,prd,twocolumn,superscriptaddress,showpacs]{revtex4}

\usepackage[dvips]{graphicx}
\usepackage{ulem}
\usepackage[usenames]{color}
\usepackage{amsmath}
\usepackage{float}
\def\slash#1{\not\!#1}

\begin{document}


\title{Negative-parity nucleon excited state in nuclear matter} 


\author{Keisuke Ohtani}
\email{ohtani.k@th.phys.titech.ac.jp}
\affiliation{Department of Physics, H-27, Tokyo Institute of Technology, Meguro, Tokyo 152-8551, Japan}

\author{Philipp Gubler}
\affiliation{Institute of Physics and Applied Physics, Yonsei University, Seoul 120-749, Korea}

\author{Makoto Oka}
\affiliation{Department of Physics, H-27, Tokyo Institute of Technology, Meguro, Tokyo 152-8551, Japan}
\affiliation{Advanced Science Research Center, Japan Atomic Energy Agency, Tokai, Ibaraki 319-1195, Japan}

\date{\today}
\begin{abstract}
Spectral functions of the nucleon and its negative parity excited state in nuclear matter are studied using QCD sum rules and the maximum entropy method (MEM). 
It is found that in-medium modifications of the spectral functions are attributed mainly to density dependencies of the $\langle \overline{q}q \rangle $ 
and $\langle q^{\dagger}q \rangle $ condensates. 
The MEM reproduces the lowest-energy peaks of both the positive and negative parity nucleon states at finite density 
up to $\rho \sim \rho_N$ (normal nuclear matter density). 
As the density grows, the residue of the nucleon ground state decreases gradually while the residue of the lowest negative parity excited state increases slightly. 
On the other hand, the positions of the peaks, which correspond to the total energies of these states, 
are almost density independent for both parity states. 
The density dependencies of the effective masses and vector self-energies are also extracted 
by assuming phenomenological mean-field type propagators for the peak states. 
We find that, as the density increases, the nucleon effective mass decreases while the vector self-energy increases. 
The density dependence of these quantities for the negative parity state on the other hand turns out to be relatively weak. 
\end{abstract}

\pacs{12.38.Lg, 14.20.Dh, 14.20.Gk}

\maketitle

\section{Introduction \label{sec:introduction}}
The question of how nucleons behave in dense matter is of great importance both from the 
point of view of nuclear physics and QCD. 
In particular, the role played by the partial restoration of chiral symmetry in nuclear matter 
and its influence on properties of the nucleonic ground and excited states has attracted continued interest. 
In this context, it is especially worth mentioning the potential medium 
modifications of the negative parity nucleon state, which are interesting from the viewpoint of 
the chiral symmetry and $\eta$ mesic nuclei. 
Considering the relation between chiral symmetry and the  spectral functions of chiral partners, 
the symmetry requires their spectral functions to be degenerate if chiral symmetry is restored. 
Chiral partners among hadronic states as well as hadronic spectral functions have been discussed already a long 
time ago \cite{Weinberg:1969hw}. 
Assuming that the chiral partner of the positive parity nucleon ground state N(939) is the lowest lying negative parity state N(1535), 
the relation between the restoration of chiral symmetry and the modifications of 
N(939) and N(1535) has been investigated within effective models such as linear sigma models \cite{Detar:1988kn,Jido:1998av}. 
These studies show that two assignments, namely the naive and mirror assignments, of the chiral transformation to the chiral partners 
lead to different characteristic modifications of the physical nucleon states at finite density. 

Potentially $\eta$ nuclear systems, so-called $\eta$-mesic nuclei, 
were first investigated by Haider and Liu \cite{Haider:1986sa}. 
The formation of $\eta$-mesic nuclei is strongly related to 
in-medium modifications of N(939) and N(1535) since the $\eta$N system strongly couples to N(1535) 
and its threshold is close to the mass of N(1535). 
Such nuclei have been studied both in theoretical \cite{Chiang:1990ft,Jido:2002yb,Nagahiro:2008rj,Friedman:2013zfa} 
and experimental approaches \cite{Chrien:1988gn,Pfeiffer:2003zd,Budzanowski:2008fr}. 
The studies of meson-nucleus bound systems are interesting because 
the hadron properties at finite density, which are related to the restoration of 
spontaneous breaking of the chiral symmetry, can be investigated in laboratories. 

In this paper, we study the spectral functions of both the positive and negative parity nucleons in nuclear matter 
using QCD sum rules. 
This method was initially developed and applied to the investigation of the meson properties in vacuum 
by Shifman \textit{et al.} \cite{Shifman:1978bx,Shifman:1978by}. 
It was subsequently used to study baryonic channels by Ioffe \cite{Ioffe:1981kw}. 
Especially for the nucleon, the analyses were thereafter continuously improved over the years by including higher order terms in the perturbative
Wilson coefficients \cite{Krasnikov:1982ea,Chung:1984gr,Jamin:1987gq,Ovchinnikov:1991mu,Shiomi:1995nf,Sadovnikova:2005ye} or non-perturbative power corrections
\cite{Belyaev:1982sa,Chung:1984gr,Leinweber:1989hh}. 
Additionally, it was pointed out that the nucleon operator couples to both positive and negative parity states \cite{Chung:1981cc}. 
The combined contributions of these states make the analysis complicated and especially spoil the result of the negative parity states.
This difficulty can be remedied by the methods of parity projection, which were proposed by 
Jido \textit{et al}. \cite{Jido:1996ia} and Kondo \textit{et al} \cite{Kondo:2005ur}. 
In these studies, the $\alpha_s$ corrections which are large for the nucleon channel were not considered. 
To include these $\alpha_s$ corrections in the parity projected sum rules, 
the present authors have improved the parity projection for baryonic QCD sum rules and studied the masses of both positive parity and negative parity 
nucleon states in vacuum \cite{Ohtani:2012ps}. 

QCD sum rules also have been used to investigate hadron properties in nuclear matter \cite{Hatsuda:1990zj,Hatsuda:1991ez,Jeong:2016qlk}. 
The generalization of nucleonic QCD sum rules in nuclear matter was 
first proposed by Drukarev \cite{Drukarev:1988kd}. 
Since then, many studies have been carried out for 
the medium modifications of its energy, effective mass and vector self-energy, which characterize 
properties of the nucleon in nuclear matter \cite{Cohen:1994wm,Drukarev:2010xv}. 
However, all previous studies have so far focused only on the positive parity state, N(939). 

In this work, we apply 
the parity-projected nucleon QCD sum rule with the phase-rotated Gaussian kernel, already used for the vacuum previously \cite{Ohtani:2012ps}, to 
the analyses in nuclear matter. 
Properties of the nucleon and its negative parity excited state are extracted from the sum rules with the help of the maximum entropy method (MEM). 
The MEM analysis combined with QCD sum rules can provide the most probable spectral function without any 
strong constraint on its form and has so far been successfully applied to the $\rho$ meson vacuum channel \cite{Gubler:2010cf}, 
the nucleon vacuum channel \cite{Ohtani:2011zz,Ohtani:2012ps} 
and others \cite{Gubler:2011ua,Araki:2014qya,Gubler:2014pta}. 
Assuming the peaks in the spectral functions to be described by in-medium nucleon propagators, 
we furthermore investigate the density dependence of the effective masses and vector self-energies of both the nucleon ground state and its negative parity 
first excited state. 

The paper is organized as follows. 
In Sec.\ \ref{Sec: OPE}, we construct the parity-projected in-medium nucleon QCD sum rules 
and discuss the behavior of the resulting equations. 
The results of the analyses are summarized in Sec.\ \ref{Sec: result} where the density dependence of the 
spectral functions of both positive and negative parity states, the effective masses and the vector self-energies are presented. 
Next, effects of the uncertainties of the condensates and their in-medium behavior on the results are studied in Sec.\ \ref{sec: Discussion}. 
We additionally discuss the validity of the parity projection at finite spatial momentum in the same section. 
Summary and conclusions are given in Sec.\ \ref{Sec: summary}.
 
\section{Parity-projected nucleon QCD sum rule in nuclear matter \label{Sec: OPE}}
\subsection{Parity projection of nucleon QCD sum rules in nuclear matter}
The parity-projected QCD sum rules are constructed from the ``forward-time'' correlation function \cite{Jido:1996ia}: 
\begin{equation}
\begin{split}
\Pi_m(q_0, |\vec{q}|) &= i \int d^{4}x e^{iqx} \theta(x_0) \\ 
                            & \hspace{0.5cm} \times \langle \Psi_0 (\rho, u^{\mu}) |T[\eta(x)\overline{\eta}(0)]| \Psi _0 (\rho, u^{\mu}) \rangle, 
\end{split}
\label{eq:forward_m2}
\end{equation}
where $\eta (x)$ is the nucleon interpolating field and 
$|\Psi_0 (\rho, u^{\mu})\rangle $ represents the ground state of nuclear matter, which is characterized 
by its velocity $u^{\mu}$ and the nucleon density $\rho$. 
We assume that $|\Psi_0 (\rho, u^{\mu})\rangle $ is invariant under parity and time reversal transformations. 
In the rest frame of nuclear matter, the velocity is given by $u^{\mu} = (1,\vec{0})$. 
Note that in Ref.\,\cite{Jido:1996ia}, this correlator was called the ``old-fashioned'' correlator. 
The essential difference from the time-ordered correlation function is the insertion of
the Heaviside step-function $\theta (x_0)$ before carrying out the Fourier transform. 
This correlator contains contributions only from states which propagate forward in time. 
With the help of the Lorentz covariance, parity invariance and time reversal invariance of the nuclear matter ground state, 
the correlation function can be decomposed into three components \cite{Cohen:1994wm}: 
\begin{equation}
\begin{split}
\Pi _m(q_0, |\vec{q}|) &=  q\hspace{-.50em}/ \Pi_{m1}(q_0, |\vec{q}|) + \Pi_{m2}(q_0, |\vec{q}|) \\
                            &\hspace{2.4cm}+  u\hspace{-.50em}/  \Pi_{m3}(q_0, |\vec{q}|).
\label{eq:correlator dec}
\end{split}
\end{equation}
The scalar functions $\Pi_{m1}$, $\Pi_{m2}$ and $\Pi_{m3}$ depend on two scalar variables $q^2$ and $q\cdot u$. 
In what follows, 
we denote ($q^2$, $q\cdot u$) as $(q_0, |\vec{q}|)$ since we will only work in the rest frame of nuclear matter. 
Note that $\Pi_{m1}$, $\Pi_{m2}$ and $\Pi_{m3}$ contain information about the in-medium properties of both positive and
negative parity states as replacing the operator $\eta(x) \rightarrow \gamma_5 \eta (x)$ only changes the sign of $\Pi_{m2}$. 

To separate these positive and negative parity contributions, we multiply the parity projection operators $\mathrm{P}^{\pm}=\frac{\gamma_0 \pm1}{2}$ 
to the correlator, take the trace over the spinor index 
and thus obtain the parity projected correlation functions: 
\begin{equation}
\begin{split}
\Pi _{m}^{+}(q_0, |\vec{q}|) &\equiv q_0 \Pi_{m1}(q_0, |\vec{q}|) +\Pi_{m2}(q_0, |\vec{q}|) \\
                                   &\hspace{2.58cm} + u_0 \Pi_{m3}(q_0, |\vec{q}|)\\
\Pi _{m}^{-}(q_0, |\vec{q}|) &\equiv q_0 \Pi_{m1}(q_0, |\vec{q}|) -\Pi_{m2}(q_0, |\vec{q}|) \\
                                   &\hspace{2.58cm} + u_0 \Pi_{m3}(q_0, |\vec{q}|). 
\end{split}
\label{eq:ppmfinal}
\end{equation}
Note that the parity projection can be carried out in accordance with that in vacuum 
because it is based on the invariance of the ground state of nuclear matter under parity transformation. 

QCD sum rules are relations between correlators computed in different regions of $q_0$. Specifically, 
$\Pi^{\pm}_{m\mathrm{OPE}}$ which is calculated at a large $-q_0^2$ by the operator product expansion (OPE) and the spectral function, 
$\rho_{m}^{\pm} \equiv \frac{1}{\pi} \mathrm{Im}[\Pi_m ^{\pm}]$ at $q_0>0$ can be related. 
Making use of the analyticity of the correlation function, one can construct the parity projected QCD sum rules: 
\begin{equation}
\begin{split}
&\int _{-\infty}^{\infty} \mathrm{Im}[\Pi ^{\pm}_{m\mathrm{OPE}} (q_0, |\vec{q}|)]  W(q_0 ) dq_0 \\
&\hspace{3cm}= \pi \int _{0} ^{\infty} \rho^{\pm}_{m} (q_0, |\vec{q}|) W( q_0 ) dq_0. 
\label{Eq: dispertion relation}
\end{split}
\end{equation}
Here we have introduced a weighting function $W( q_0 )$, which is real at real $q_0$ and 
analytic in the upper half of the imaginary plane of $q_0$. 
The details of the derivation of Eq.\,(\ref{Eq: dispertion relation}) are discussed in \cite{Ohtani:2012ps}. 

\subsection{Operator product expansion in nuclear matter\label{subsec: construction}}
In this subsection, we provide the explicit form of $\Pi^{\pm}_{m\mathrm{OPE}}$ including all known 
$\alpha_{s}$ corrections. 
For the nucleon, there are two independent local interpolating operators: 
\begin{equation}
\eta _{1}(x)=\epsilon ^{abc} \bigl[ u^{Ta}(x)C\gamma_{5} d^{b}(x) \bigr] u^{c}(x),
\label{eq:eta1}
\end{equation}
\begin{equation}
\eta _{2}(x)=\epsilon ^{abc} \bigl[ u^{Ta}(x)Cd^{b}(x) \bigr] \gamma_{5} u^{c}(x). 
\label{eq:eta2}
\end{equation}
Here, $a$, $b$ and $c$ are color indices, $C= i\gamma_0 \gamma_2$ stands for the charge conjugation matrix, while the spinor indices are omitted for simplicity. 
A general interpolating field can be expressed as 
\begin{equation}
\eta(x)= \eta_{1}(x)+\beta\eta_{2}(x),
\label{eq:interpolating field}
\end{equation}
where $\beta$ is a real parameter. 
The choice $\beta=-1$ is called the Ioffe current, which is widely used in sum rule analyses studying the nucleon ground state. 
It is straightforward to obtain the imaginary part of the forward-time correlator of Eq.(\ref{eq:forward_m2}) from 
the time ordered correlator given in the literature \cite{Ovchinnikov:1991mu,Ohtani:2012ps,Groote:2008hz}. 
The explicit expressions are given as 
\begin{equation}
\begin{split}
&\frac{1}{\pi}\mathrm{Im}\left [q_0 \Pi _{m1\mathrm{OPE}}(q_{0}, |\vec{q}|) \right ] \\
& \quad =  \frac{C_1}{2^{11}\pi^4} \left[1 + \frac{\alpha_s}{\pi} \left ( \frac{71}{12} - \ln (\frac{q^2}{\mu ^{2}}) \right ) \right ]  \\
       & \hspace{1cm} \times q_{0}(q^2)^2  \theta (q_{0}-|\vec{q}|)  \\
       & \quad \quad + \frac{C_1}{2^{10}\pi^2} \langle \frac{\alpha _s}{\pi }G^{2}\rangle _{m}  q_0\ \theta (q_{0}-|\vec{q}|) \\
       & \quad \quad -\frac{C_1}{2^{5} 3^2\pi^2} \langle q^\dagger iD_0q\rangle _{m} \\ 
       & \hspace{1cm} \times \bigl[ 5q_0 \theta(q_0-|\vec{q}|) -4 |\vec q|^2 \delta(q_0-|\vec{q}|) \bigr ]  \\     
       & \quad \quad -\frac{C_1}{2^{9} 3^2\pi^2} \langle \frac{\alpha _s}{\pi } (E^2+B^2) \rangle _{m} \\
       & \hspace{1cm} \times \bigl[ q_0 \theta(q_0-|\vec{q}|) -2 |\vec q|^2 \delta(q_0-|\vec{q}|) \bigr ]  \\ 
       & \quad \quad +\frac{1}{2^4 3} \left (  C_2 + \frac{\alpha_s}{\pi} C_3 \right ) \langle \overline{q}q\rangle ^2 _{m}  \delta (q_{0}-|\vec{q}|)  \\
       & \quad \quad -\frac{C_4}{2^3 3 \pi} \frac{\alpha_s}{\pi} \langle \overline{q}q\rangle ^2 _{m}  \mathrm{Im}\biggl [ \ln (2|\vec{q}|) \frac{|\vec{q}|}{(q_0+i\epsilon )^2-|\vec{q}|^2}  \\
       & \hspace{1cm} + \ln \bigl (|\vec{q}|-(q_0+i\epsilon )\bigr ) \frac{q_0}{(q_0+i\epsilon )^2-|\vec{q}|^2} \biggr ] \\ 
       & \quad \quad +  \frac{C_1}{2^{4} 3} \langle q^{\dagger} q\rangle _{m}^2 \delta (q_{0}-|\vec{q}|) \\
       & \quad \quad - \frac{C_1}{2^5 3 \pi^2 } \langle q^{\dagger} q\rangle _{m} q_0^2 \ \theta(q_0-|\vec{q}|) \\
       & \quad \quad - \frac{1}{2^5 3 \pi^2 } \left (C_5 - C_1 \ln (\frac{q^2}{\mu ^{2}}) \right ) \\ 
       & \hspace{1cm} \times \frac{\alpha_s}{\pi} \langle q^{\dagger} q\rangle _{m} q_0^2 \ \theta(q_0-|\vec{q}|) \\       
       & \quad \quad - \frac{C_6}{2^5 3^2 \pi^2} \langle q^\dagger g\sigma \cdot Gq\rangle  _{m} \ \frac{|\vec q|}{2} \delta (q_0-|\vec q|)  \\
       & \quad \quad - \frac{C_1}{2^4 3 \pi^2} \left [ \langle q^\dagger i D_0 i D_0 q\rangle  _{m} +\frac{1}{12}\langle q^\dagger g\sigma \cdot Gq\rangle  _{m} \right ] \\ 
       & \hspace{1cm} \times \biggl (  -2 |\vec{q}|\delta (q_0-|\vec{q}|) + 2|\vec{q}|^4 \\
       & \hspace{1cm} \times  \mathrm{Im} \bigl [ \frac{1}{4\pi |\vec q|^2 -i\epsilon } \cdot \frac{1}{(q_{0}-|\vec q|+i\epsilon )^2} \bigr ] \biggr ) 
\label{eq:Pi1_for_s}
\end{split}
\end{equation}
\begin{equation}
\begin{split}
& \frac{1}{\pi} \mathrm{Im}\left [\Pi _{m2\mathrm{OPE}}(q_{0},|\vec{q}|) \right ] \\ 
&\quad = -\frac{1}{2^6 \pi^2} \left ( C_2 +C_7 \frac{\alpha_s}{\pi} \right ) \langle \overline{q}q\rangle _{m} q^{2} \theta (q_{0}-|\vec{q}|)  \\ 
& \quad \quad + \frac{3 C_8}{2^6 \pi^2} \langle \overline{q}g\sigma \cdot Gq\rangle  _{m} \ \theta (q_{0}-|\vec{q}|)  \\
& \quad \quad -\frac{C_9}{6 \pi^2} \left[ \langle \overline{q} i D_0 i D_0 q\rangle  _{m} +\frac{1}{8}\langle \overline{q} g\sigma \cdot Gq\rangle _{m} \right ] \\
& \hspace{1cm} \times  \frac{|\vec{q}|}{2}\delta (q_{0}-|\vec{q}|) \\                    
& \quad \quad +\frac{C_2}{2^5 \pi^2}\langle \overline{q}iD_{0}q \rangle _{m} q_0 \theta (q_0-|\vec{q}|) \\
& \quad \quad + \frac{C_2}{2^3 3} \langle \overline{q}q\rangle _{m} \langle q^{\dagger}q\rangle _{m}\ \delta (q_0-|\vec{q}|) 
\label{eq:Pi2_for_s}
\end{split}
\end{equation} 
\begin{equation}
\begin{split}
&\frac{1}{\pi} \mathrm{Im} \left [\Pi _{m\mathrm{3OPE}}(q_{0},|\vec{q}|) \right ] \\ 
&\quad = \frac{5 C_1}{2^3 3^2 \pi^2} \langle q^{\dagger}iD_{0}q \rangle _{m}  q_0 \theta(q_0 -|\vec{q}|) \\  
   &\quad \quad +\frac{C_1}{2^7 3^2 \pi^2} \langle \frac{\alpha_s}{\pi} (E^2+B^2)\rangle _{m}  q_0 \theta(q_0 -|\vec{q}|) \\  
   &\quad \quad + \frac{C_1}{2^3 3} \langle q^{\dagger} q \rangle ^2  _{m} \ \delta (q_0-|\vec{q}|) \\
   &\quad \quad - \frac{C_1}{2^4 3 \pi^2} \langle q^{\dagger} q \rangle _{m}  (q_0^2-|\vec{q}|^2) \ \theta (q_0 -|\vec{q}|) \\ 
   &\quad \quad - \frac{1}{2^4 3 \pi^2} \left (C_{10} - C_1 \ln (\frac{q^2}{\mu ^{2}}) \right ) \frac{\alpha_s}{\pi} \langle q^{\dagger} q \rangle _{m}  \\
   & \hspace{1cm} \times (q_0^2-|\vec{q}|^2) \ \theta (q_0 -|\vec{q}|) \\ 
   &\quad  \quad + \frac{C_6}{2^6 3 \pi^2}  \langle q^{\dagger}g\sigma \cdot Gq\rangle  _{m} \theta(q_0 -|\vec{q}|) \\
   &\quad  \quad - \frac{C_1}{2^3 \pi^2} \left [ \langle q^\dagger i D_0 i D_0 q\rangle  _{m}  +\frac{1}{12}\langle q^\dagger g\sigma \cdot Gq\rangle  _{m} \right ] \\
   &\hspace{1cm} \times \frac{|\vec q|}{2} \delta ((q_{0}-|\vec q|), 
\label{eq:Piu_for_s}
\end{split}
\end{equation}
where $q^2=q_0^2- \vec{q}^2$ and the coefficients $C_i$ are defined as
\begin{equation}
\begin{split}
C_1 &= 5 + 2\beta + 5\beta^2 \\
C_2 &= 7-2\beta -5\beta^2 \\
C_3 &= \frac{325}{18} + \frac{448}{9} \beta  + \frac{511}{18} \beta ^2 \\
C_4 &= \frac{47}{3}  - \frac{10}{3} \beta - \frac{61}{3} \beta ^2 \\
C_5 &= \frac{49}{3}  + \frac{14}{3} \beta + \frac{49}{3} \beta ^2 \\
C_6 &= 7 + 10\beta + 7\beta^2 \\
C_7 &= \frac{15}{2} -3 \beta - \frac{9}{2} \beta ^2  \\
C_8 &= 1-\beta^2 \\
C_9 &=2-\beta -\beta^2 \\
C_{10} &= \frac{211}{12} + \frac{31}{6} \beta + \ \frac{211}{12} \beta ^2. 
\end{split}
\end{equation}
The matrix elements $\langle \mathcal{O} \rangle _m$ stand for the 
expectation value of operators $\mathcal{O}$ in nuclear matter. 

\subsection{QCD condensates at finite nucleon density} 
The correlation functions are characterized by in-medium QCD condensates. 
While the value of the vector quark condensate $\langle q^{\dagger} q \rangle _{m}$ at density $\rho$ is $\frac{3}{2} \rho$ exactly, 
the other condensates are not precisely determined and their density dependences may be more complicated. 
In this paper, we estimate their values in the linear density approximation, which is valid at sufficiently low density \cite{Cohen:1991nk,Drukarev:1988kd}. 
The in-medium condensates $\langle \mathcal{O} \rangle _{m}$ are in this approximation expressed 
as $\langle \mathcal{O} \rangle _{m} =\langle \mathcal{O} \rangle_0 + \rho \langle \mathcal{O} \rangle _N $ with 
the vacuum condensates $\langle \mathcal{O} \rangle_0$ 
and the nucleon matrix elements  $\langle \mathcal{O} \rangle _N \equiv \langle N | \mathcal{O} | N \rangle$. 
Each matrix element is evaluated as follows: 
\begin{equation}
\begin{split}
\langle \overline{q}q\rangle _{m} &= \langle \overline{q}q\rangle _{0} + \rho \langle \overline{q}q\rangle _{N} \\
                                           &=\langle \overline{q}q\rangle _{0} + \rho \frac{\sigma_ N}{2m_q} \\
\langle q^{\dagger} q\rangle _{m} &= \rho \frac{3}{2} \\
\langle \frac{\alpha _s}{\pi }G^{2}\rangle _{m} &= \langle \frac{\alpha _s}{\pi }G^{2}\rangle _{0} + \rho \langle \frac{\alpha _s}{\pi }G^{2}\rangle _{N}  \\
\langle q^\dagger iD_0q\rangle _{m} &= \rho \langle q^\dagger iD_0q\rangle _{N}  =\rho \frac{3}{8} M_N A_2^q \\
\langle \frac{\alpha _s}{\pi } (E^2+B^2) \rangle _{m} &= \rho \langle \frac{\alpha _s}{\pi } (E^2+B^2) \rangle _{N} \\
                                                                     &= \rho \frac{3}{2\pi} M_N \alpha_s (\mu^2) A_2^g   \\
\langle \overline{q} i D_0 q\rangle _{m} &=  m_q \langle q^{\dagger} q \rangle _{m}  \simeq 0\\
\langle \overline{q}g\sigma \cdot Gq\rangle _{m}&=\langle \overline{q}g\sigma \cdot Gq\rangle _{0} + \rho \langle \overline{q}g\sigma \cdot Gq\rangle _{N} \\
                                                                 & \approx m_0 ^2 \langle \overline{q}q\rangle _{m}  \\
                                                                       \langle q^\dagger g\sigma \cdot Gq\rangle _{m} &=  \rho \langle q^\dagger g\sigma \cdot Gq\rangle _{N} \\
\langle q^\dagger i D_0 i D_0 q\rangle  _{m} +&\frac{1}{12}\langle q^\dagger g\sigma \cdot Gq\rangle _{m} \\
&= \bigl ( \langle q^\dagger i D_0 i D_0 q\rangle _N +\frac{1}{12}\langle q^\dagger g\sigma \cdot Gq\rangle _N \bigr ) \rho \\
&=\rho \frac{1}{4} M_N^2 A_3^q \\
\langle \overline{q} i D_0 i D_0 q\rangle _{\rho_N}+ & \frac{1}{8}\langle \overline{q} g\sigma \cdot Gq\rangle _{m} \\
&= \bigl (\langle \overline{q} i D_0 i D_0 q\rangle _{N}+\frac{1}{8}\langle \overline{q} g\sigma \cdot Gq\rangle _{N} \bigr ) \rho \\
&= \rho  \frac{3}{4} M_N^2 e_2,
\end{split}
\label{eq:condensates1}
\end{equation}
where $E$ and $B$ are the color electric and color magnetic fields, respectively. 
$\langle \overline{q} q \rangle$ denotes the averages over up and down quarks, 
$\frac{1}{2} \left (\langle \overline{u} u \rangle + \langle \overline{d} d \rangle \right )$. 

The quantities $A_2^q$, $A_2^g$, $A_3^q$, $e_2$ can be expressed as 
moments of the parton distribution functions \cite{Cohen:1994wm}. 
The values of the parameters appearing in Eq.\,(\ref{eq:condensates1}) is given in Table \ref{tab:the values of the condensate in nuclear matter}. 
The uncertainties of the values of $m_q$ and $\sigma_N$ will be discussed in Sec.\ \ref{sec: Discussion}.  
Note that the higher-order density terms of the chiral condensate $\langle \overline{q} q \rangle$ have been computed using chiral perturbation 
theory \cite{Kaiser:2007nv,Goda:2013bka}. 
These contributions are however small up to the normal nuclear mater density and thus 
we do not take them into account in this study. 

\begin{table}[t!]
\begin{center}
\begin{tabular}{|c|c|}
\hline 
parameters & values \\ \hline
$\langle \overline{q}q\rangle _{0}$  & $-(0.246\pm 0.002 \mathrm{GeV})^3$ \cite{Borsanyi:2012zv} \\ \hline
$m_q$  & $4.725\mathrm{MeV}$ \cite{Agashe:2014kda}  \\ \hline
$\sigma _N$  & $45 \mathrm{MeV}$   \\ \hline
$\langle q^{\dagger} q\rangle _{m}$ &  $\rho \frac{3}{2}$ \\ \hline
$\langle \frac{\alpha _s}{\pi }G^{2}\rangle _{0}$ & $0.012 \pm 0.0036 \mathrm{GeV}^4$  \cite{Colangelo:2000dp} \\ \hline 
$\langle \frac{\alpha _s}{\pi }G^{2}\rangle _{N}$ & $-0.65 \pm 0.15\mathrm{GeV}$ \cite{Jin:1993up} \\ \hline 
$A^q_2$ & $0.62 \pm 0.06$ \cite{Martin:2009iq} \\ \hline
$A^g_2$ & $0.359 \pm 0.146$ \cite{Martin:2009iq} \\ \hline
$A^q_3$ & $0.15 \pm 0.02$ \cite{Martin:2009iq} \\ \hline
$e_2$ & $0.017\pm0.047$ \cite{Gubler:2015uza} \\ \hline
$m_0^2$ & $0.8 \pm 0.2 \mathrm{GeV}^2$ \cite{Colangelo:2000dp} \\ \hline
$\langle q^\dagger g\sigma \cdot Gq\rangle _{N}$ & $-0.33 \mathrm{GeV}^2$ \cite{Jin:1993up} \\ 
\hline
\end{tabular}
\caption{Values of parameters appearing in Eq.\,(\ref{eq:condensates1}). }
\label{tab:the values of the condensate in nuclear matter}
\end{center}
\end{table}

\subsection{Phase-rotated Gaussian QCD sum rules}
To explicitly compute both the left and the right hand sides of Eq.\,(\ref{Eq: dispertion relation}), 
we have to specify the kernel $W(q_0)$. 
In a previous study, in which the nucleon properties in vacuum were investigated \cite{Ohtani:2012ps}, we tested several kinds of 
kernels such as the Borel and Gaussian kernels 
and found that the phase-rotated Gaussian kernel is most suitable for studying the nucleon ground state and its negative parity excitation. 
As it was pointed out in Ref.\,\cite{Ohtani:2012ps}, choosing an appropriate phase parameter $\theta$, the kernel 
improves the convergence of the OPE and at the same time suppresses the $\alpha_s$ corrections. 
Moreover, the four quark condensate contributions are suppressed with this kernel, 
and therefore the uncertainties caused by the four quark condensates, whose values are only weakly constrained,  
will not seriously affect the results of the QCD sum rule analysis. 

We will later carry out the sum rule analysis for the nucleon at rest relative to nuclear matter and also at the Fermi surface. 
There is no guarantee that the above desirable features are kept 
when investigating the in-medium nucleon properties at finite $|\vec{q}|$. 
We, in fact, find that the suppression of the contributions from the $\alpha_s$ corrections becomes less effective 
as $|\vec{q}|$ increases.  
Therefore, we improve the phase rotated kernel $W(q_0,|\vec{q}|)$ as 
\begin{equation}
\begin{split}
&W(s,\tau, \theta, q_0, |\vec{q}|) =\\
&\frac{1}{\sqrt{4\pi\tau}} \frac{1}{2} \Biggl[ (q_0-|\vec{q}|) e^{-2i\theta} \exp \Bigl(  -\frac{(q^2 e^{-2i\theta} - s)^2}{4\tau} \Bigr) \\
&\hspace{1cm} + (q_0-|\vec{q}|) e^{2i\theta} \exp \Bigl(  -\frac{(q^2 e^{2i\theta} - s)^2}{4\tau} \Bigr)\Biggr].
\end{split}
\label{eq:phaserokernel2}
\end{equation}
 \begin{figure*}[!tbp]
 \begin{center}
  \includegraphics[width=11.5cm]{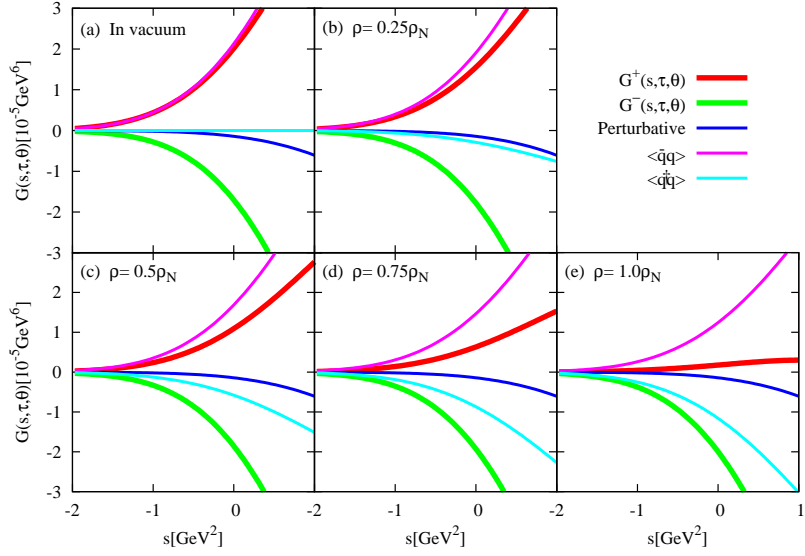}
 \end{center}
 \caption{
 The density dependence of $G^{\pm}_{m\mathrm{OPE}}(s, \tau, \theta)$. The perturbative, chiral condensate $\langle \overline{q}q \rangle$, 
vector quark condensate $\langle q^{\dagger}q \rangle$ terms and the total are shown at $\tau=0.5 \mathrm{GeV}^4$, $\beta=-0.9$, $\theta = 0.108\pi$ and 
$|\vec{q}| = 0 $. 
The $\langle q^{\dagger}q \rangle$ term stands for the sum of the terms proportional to the condensate $\langle q^{\dagger}q \rangle$ 
in $G_{m1\mathrm{OPE}}(s,\tau, \theta)$ and $G_{m3\mathrm{OPE}}(s,\tau, \theta)$. 
}
 \label{fig:OPE density}
\end{figure*}
$s$, $\tau$ and $\theta$ are parameters in our QCD sum rule and whose analyzed parameter regions will be discussed in the next section. 
For $|\vec{q}|=0$, the above kernel is equivalent to the one used previously in Ref.\,\cite{Ohtani:2012ps}. 

Substituting Eqs.(\ref{eq:Pi1_for_s}-\ref{eq:Piu_for_s}) and (\ref{eq:phaserokernel2}) into Eq.(\ref{Eq: dispertion relation}), we finally obtain 
the parity-projected nucleon QCD sum rules as
\begin{equation}
\begin{split}
&G^{\pm}_{m\mathrm{OPE}}(s,\tau, \theta) \\
&\equiv \int_{-\infty }^{\infty}\frac{1}{\pi}\mathrm{Im}\Bigl [\Pi^{\ \pm }_{m\mathrm{OPE}}(q_{0},|\vec{q}|)\Bigr ] W(s,\tau, \theta, q_{0}, |\vec{q}|) dq_{0} \\
&= \mathrm{G_{m1\mathrm{OPE}}}(s,\tau, \theta) \pm \mathrm{G_{m2\mathrm{OPE}}}(s,\tau, \theta) + \mathrm{G_{m3\mathrm{OPE}}}(s,\tau, \theta) \\
&= \int_{0}^{\infty} \rho ^{\pm}_{m}(q_{0}) W(s,\tau, \theta, q_{0}, |\vec{q}|) dq_{0}.
\label{eq:Gmatter}
\end{split}
\end{equation}
Here, $\mathrm{G_{mi\mathrm{OPE}}}(s,\tau, \theta) \ (i=1,2,3)$ are defined as 
\begin{equation}
\begin{split}
G_{m1\mathrm{OPE}}(s,\tau, \theta) &= \int _{-\infty}^{\infty } \mathrm{Im}\biggl [\frac{q_0 \Pi _{m1\mathrm{OPE}}(q_{0},|\vec{q}|)}{\pi} \biggr ] \\ 
                                                  & \hspace{1cm} \times W(s,\tau, \theta, q_{0},|\vec{q}|) dq_{0} \\
G_{m2\mathrm{OPE}}(s,\tau, \theta) &= \int _{-\infty}^{\infty } \mathrm{Im}\biggl [\frac{\Pi _{m2\mathrm{OPE}}(q_{0},|\vec{q}|)}{\pi} \biggr ]  \\
                                                  & \hspace{1cm} \times W(s,\tau, \theta, q_{0},|\vec{q}|) dq_{0} \\
G_{m3\mathrm{OPE}}(s,\tau, \theta) &= \int _{-\infty}^{\infty } \mathrm{Im}\biggl [\frac{\Pi _{m\mathrm{u}\mathrm{OPE}}(q_{0},|\vec{q}|)}{\pi} \biggr ]  \\
                                                  & \hspace{1cm} \times W(s,\tau, \theta, q_{0},|\vec{q}|) dq_{0}.  
\label{eq:Gmatter12u}
\end{split}
\end{equation}
The functions $G^{\pm}_{m\mathrm{OPE}}(s,\tau, \theta)$ 
are shown in Fig.\,\ref{fig:OPE density} at $\tau = 0.5 [\mathrm{GeV}^4]$, $\theta = 0.108 \pi$ and $|\vec{q}| =0 $ for various densities. 
The qualitative behavior at finite spatial momentum is similar to that at $|\vec{q}|=0$. 
In this figure, also shown are the perturbative, chiral condensate $\langle \overline{q}q \rangle$ 
and vector quark condensate $\langle q^{\dagger }q \rangle$ terms, which are dominant. 
From Eq.(\ref{eq:Gmatter}) and Fig. \ref{fig:OPE density}, one sees that the chiral condensate term dominates in vacuum and is 
responsible for the difference between $G^{+}_{m\mathrm{OPE}}$ and $G^{-}_{m\mathrm{OPE}}$. 
This observation shows clearly that the difference between the positive and negative-parity spectral functions 
is caused by the emergence of the chiral condensate $\langle \overline{q}q \rangle$. 
We also find that, as the density increases, $G^{+}_{m\mathrm{OPE}}$ becomes small 
due to the decrease of the absolute value of $\langle \overline{q}q \rangle$ and the increase of the vector quark condensate. 
On the other hand, $G^{-}_{m\mathrm{OPE}}$ shows no significant change since the modifications of the $\langle \overline{q}q \rangle$ and 
$\langle q^{\dagger} q \rangle$ condensates cancel each other out to a large degree. 
\begin{figure*}[!tbp]
\begin{center}
  \includegraphics[width=9.cm]{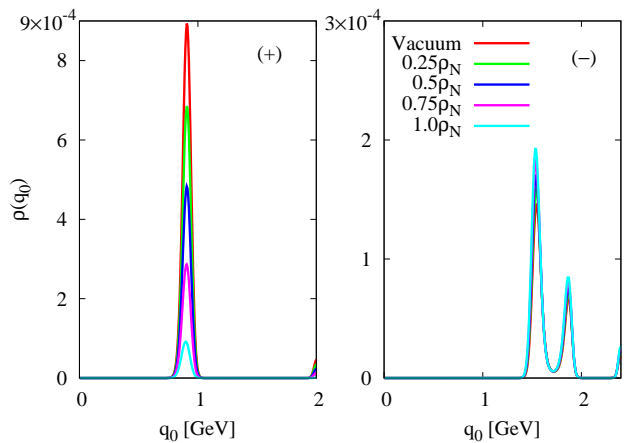}
 \end{center}
 \caption{The positive (left) and negative (right) parity spectral functions extracted from $G^{\pm}_{m\mathrm{OPE}} (s,\tau, \theta )$ 
of Eq.(\ref{eq:Gmatter}) by MEM. 
The red, green, blue, magenta and light blue lines correspond to the spectral functions at the density $0.0\rho_N$, $0.25\rho_N$, $0.5\rho_N$, $0.75\rho_N$ 
and $1.0\rho_N$, respectively. Here $\rho_N$ denotes the normal nuclear matter density.
}
 \label{fig:SPF100restframe}
\end{figure*}

\section{Numerical analysis of the sum rules \label{Sec: result}}
\subsection{Spectral functions of the positive and negative-parity states}
We first discuss the parameter regions of $\tau$, $s$, $\theta$ and $\beta$ used for the analyses of this work. 
Considering the form of $W(s,\tau, \theta, q_0, |\vec{q}| )$ in Eq.(\ref{eq:phaserokernel2}), we expect that 
for small $\tau$ values, $G^{\pm}_{m}(s, \tau, \theta)$ will retain traces of the 
peak structures of the spectral function, while at large $\tau$, it will be dominated by 
continuum contributions. 
This is because $\tau$ represents the typical energy scale over which the kernel $W(s,\tau, \theta, q_0, |\vec{q}| )$ averages 
the spectral function. 
We therefore use several values of $\tau$ ($\tau = 0.5, 0.75, 1.0, 1.25, 1.5, 1.75, 2.0 \ \mathrm{GeV}^4$) simultaneously 
and determine 
the corresponding parameter region of $s$ for each $\tau$. 
The minimum values of $s$ at fixed $\tau$ are determined 
based on the criterion that the ratio of the highest dimensional OPE term to the total $G_{m\mathrm{OPE}}^{\pm} (s, \tau, \theta)$
is less than $0.25$. 
The maximum values of $s$ are chosen to satisfy the condition that  
the second node of the kernel as a function of $q_0$ is less than 2.0 GeV because it 
is difficult to extract information in the $q_0$ region above this second node 
due to the suppression and fast oscillation of the kernel. 
The specific values of the minimum and maximum $s$ for each $\tau$ are shown in 
Table\,\ref{tab:s_{min} at beta=-0.9 only four quark}. 
The values of $\theta$ and $\beta$ are set to $0.108 \pi$ and $-0.9$ to 
suppress the effects of higher order $\alpha_s$ corrections and uncertainties of condensates, respectively. 
For more details about the parameter determination, we refer the reader to Ref.\,\cite{Ohtani:2012ps}. 

We apply the maximum entropy method (MEM) to the OPE data of Eq.(\ref{eq:Gmatter}) 
and extract the spectral functions of both positive and negative parity states. 
The advantage of this method is that the most probable spectral function can be obtained without assuming its 
specific form such as the ``pole + continuum'' ansatz \cite{Gubler:2010cf}. 
In the MEM analysis, we however have to introduce the so-called default model $m(q_0 )$, which should include our prior knowledge of the spectral function. 
To correctly reflect the spectral behavior both in the high and low energy regions, we use the following default model, 
\begin{equation}
\begin{split}
m(q_{0} )&= \bar{m}(\beta )
\frac{1}{1+e^{(q_{\mathrm{th}} -q_{0} )/\delta}}, \\
\bar{m}(\beta ) &= \frac{5+2\beta +5\beta ^2 }{128(2\pi ) ^4}
\label{eq:default hybr_matter}
\end{split}
\end{equation}
where the values of $q_{\mathrm{th}}$ and $\delta$ are chosen as $3.0$ GeV and $0.1$ GeV, respectively. 
The factor $\bar{m}(\beta )$ is determined so that $m(q_0)$ agrees with 
the asymptotic behavior of the spectral function at high energy. 
For further technical details of MEM, we refer the reader to \cite{Asakawa:2000tr,PhysRevLett.65.496,Gubler:2010cf}. 
The errors of the OPE data in vacuum $\sigma (s, \tau)_{\rho=0}$ are evaluated based on the method proposed in Ref.\,\cite{Leinweber:1995fn}, 
while the errors $\sigma (s, \tau) _{\rho=\rho_N} $ in nuclear matter are determined by assuming that the relative errors are density independent, 
$\left ( \sigma (s, \tau) /G_{m\mathrm{OPE} (s, \tau)} \right )_{\rho=\rho_N} = \left ( \sigma (s, \tau)/G_{m\mathrm{OPE}} (s, \tau) \right ) _{\rho=0}$. 
\begin{table}[t!]
\begin{center}
\begin{tabular}{|c|c|c|c|c|c|c|c|}
\hline 
$\tau$                                         &0.5 &0.75 &1.0  &1.25 &1.5  &1.75 &2.0  \\ \hline
$s_{\mathrm{min}}$ of $G^{+}_{m}$ &-2.44&-3.92&-5.41&-6.90&-8.39&-9.88&-11.37 \\ \hline
$s_{\mathrm{max}}$ of $G^{+}_{m}$ &0.90&-0.10&-1.20&-2.20&-3.40&-4.50&-5.70  \\ \hline
$s_{\mathrm{min}}$ of $G^{-}_{m}$ &-1.27&-2.26&-3.27&-4.28&-5.30&-6.32&-7.35  \\ \hline
$s_{\mathrm{max}}$ of $G^{-}_{m}$ &0.90&-0.10&-1.20&-2.20&-3.40&-4.50&-5.70  \\ 
\hline
\end{tabular}
\caption{Values of $s_{\mathrm{min/max}}$ [GeV$^2$] at $\beta=-0.9$ and fixed $\tau$ [GeV$^4$].}
\label{tab:s_{min} at beta=-0.9 only four quark}
\end{center}
\end{table}

We first analyze the in-medium spectral functions of the nucleons at rest relative to nuclear matter ($\vec{q}$ =0). 
The obtained spectral functions are shown in Fig.\,\ref{fig:SPF100restframe}. 
For positive parity, the peak appears at about 910 MeV in vacuum. 
This peak corresponds to the nucleon ground state N(939). 
As the density increases, the height of the peak decreases while the peak position does not change much. 
For negative parity, two peaks appear at 1550 MeV and 1870 MeV in vacuum. 
The first peak lies close to the lowest negative parity excitation N(1535) and thus most likely corresponds to this state. 
Note, however, that it is generally difficult to disentangle two adjoining peaks in a narrow region from QCD sum rule analyses. 
It is therefore possible that the lowest peak contains contributions of the higher N(1650) state. 
The second peak is statistically less significant than the first and its position does not correspond to any known $\frac{1}{2}^{-}$ nucleon 
excited sates. 
This peak may therefore be a manifestation of the continuum. 
Consistently with the behavior of the OPE data shown in Fig.\,\ref{fig:OPE density}, the negative parity spectral function is not modified significantly 
at finite density. 
These findings indicate that the energies of the lowest lying states of both positive and negative parity 
are almost density independent while the coupling strength of the employed interpolating field to the N(939) state decreases as the density increases. 

\subsection{Estimation of self energies}
So far, we have found that the peak positions, namely the total energies of the nucleon and its negative parity excited state, 
are not sensitive to matter effects up to nuclear matter density. 
The behavior of the nucleon ground state is consistent 
with the small binding energy per nucleon of nuclear matter. 
The results for the negative parity state are on the other hand unexpected 
because one would naively anticipate 
that its peak moves towards the peak of the positive parity spectral function 
as the chiral symmetry is partially restored in the nuclear medium.

The quantum hadrodynamics (QHD) model has been successfully applied to the investigation of 
nuclei and in-medium nucleon properties \cite{Serot1,Serot2}. 
In this framework, the nearly-density-independent single particle energy of the nucleon in nuclei 
is caused by the cancellation of the scalar and vector self-energies. 
The investigation of the self-energies of the negative parity state, 
to be carried out in this subsection within our QCD sum rule approach, 
will hence be similarly helpful to comprehend its remarkable behavior. 
\begin{figure*}[!tbp]
\begin{center}
  \includegraphics[width=10.2cm]{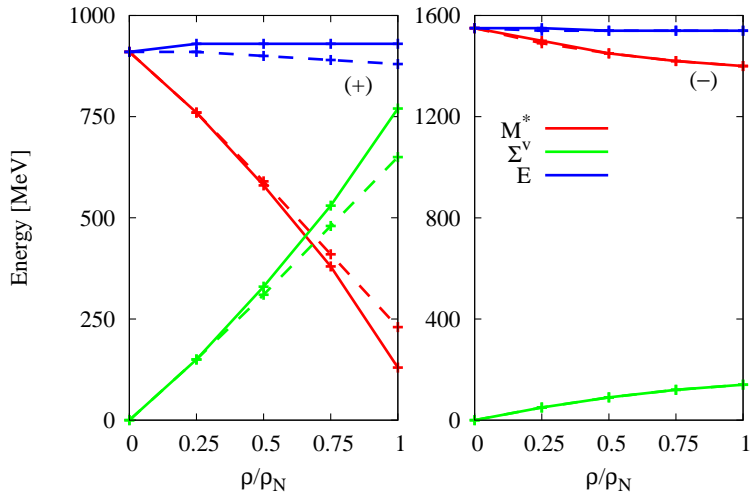}
 \end{center}
 \caption{The density dependence of the effective masses and vector self energies of 
positive (left) and negative parity (right) states. 
The red, green and blue lines correspond to the effective masses, vector self-energies and total energies, respectively. 
The dashed lines are the results in which the four quark condensate are assumed to be independent of the density 
(see \ref{subsec: in-medium four quakr condensate dependence}). 
}
 \label{fig:selfenergiesrestframe}
\end{figure*}

Consider the nucleon propagator in nuclear matter, 
\begin{equation}
\begin{split}
G(q_0,|\vec{q}|) = \frac{Z^{'}(q_0, |\vec{q}|)}{\slash q - M - \Sigma (q_0, |\vec{q}|) + i\epsilon},
\label{Eq:npropagator}
\end{split}
\end{equation}
where $\Sigma(q_0, |\vec{q}|)$ is the nucleon self-energy and $Z^{'}(q_0, |\vec{q}|)$ means the renormalization factor of 
the nucleon wave function. 
As in Eq.(\ref{eq:correlator dec}), the self-energy can be decomposed as 
\begin{equation}
\begin{split}
\Sigma (q) = \Sigma ^{s'} (q_{0}, |\vec{q}|) + \Sigma ^{v'} (q_{0}, |\vec{q}|) \slash{u} + \Sigma ^{q'} (q_{0}, |\vec{q}|) \slash{q}. 
\label{Eq:SRself}
\end{split}
\end{equation}
It turns out to be convenient to redefine the quantities $\Sigma^{q'}$ to $\Sigma ^{s'}$, $\Sigma ^{v'}$ and $Z^{'}$ as  
\begin{equation}
\begin{split}
M^{*} &\equiv \frac{M+ \Sigma ^{s'} (q_{0}, |\vec{q}|)}{1-\Sigma^{q'}} =M + \Sigma ^s (q_{0}, |\vec{q}|) ,\\
\Sigma ^v (q_{0}, |\vec{q}|) &\equiv  \frac{\Sigma ^{v'} (q_{0}, |\vec{q}|)}{1-\Sigma ^{q'}}, \\ 
Z(q_0, |\vec{q}|) &\equiv \frac{Z^{'} (q_{0}, |\vec{q}|)}{1-\Sigma^{q'}},
\end{split}
\end{equation}
after which the nucleon propagator can be described as 
\begin{equation}
\begin{split}
G(q_0, |\vec{q}|) = Z(q_0, |\vec{q}|)\frac{\slash q - \slash u \Sigma_v + M^*}{(q_0 - E +i \epsilon )(q_0 + \overline{E} - i \epsilon )}, 
\end{split}
\end{equation}
where
\begin{equation}
\begin{split}
E = \Sigma_v + \sqrt{M^{*2} + \vec{q}^2}, \ \ \overline{E} = -\Sigma_v + \sqrt{M^{*2} + \vec{q}^2}.
\end{split}
\end{equation}
Now we assume that the phenomenological side of the nucleon correlation function is constituted of several sharp (zero-width) 
positive and negative parity states and the continuum. 
Then, each scalar function of the forward-time correlation function can be expressed by 
a sum of the contributions from the individual states as   
\begin{equation}
\begin{split}
q_0 \Pi_{m1}(q_0, |\vec{q}| ) &= \sum _n |\lambda _n^+|^2 \frac{E_n^+}{2\sqrt{M^{*2}_{n+}+\vec{q}^2}}\frac{1}{q_0 - E_n^+ + i\epsilon}\\ 
 & +|\lambda _n^-|^2 \frac{E_n^-}{2\sqrt{M^{*2}_{n-}+\vec{q}^2}}\frac{1}{q_0 - E_n^- + i\epsilon} + \cdots, 
\end{split}
\label{eq:OPEm1}
\end{equation}
\begin{equation}
\begin{split}
 \Pi_{m2}(q_0,|\vec{q}|) &= \sum _n |\lambda _n^+|^2 \frac{M^*_{n+}}{2\sqrt{M^{*2}_{n+}+\vec{q}^2}}\frac{1}{q_0 - E_n^+ + i\epsilon} \\
 & -|\lambda _n^-|^2 \frac{M^{*}_{n-}}{2\sqrt{M^{*2}_{n-}+\vec{q}^2}}\frac{1}{q_0 - E_n^- + i\epsilon} + \cdots, 
\end{split}
\end{equation}
\begin{equation}
\begin{split}
 \Pi_{m3}(q_0,|\vec{q}|) &= \sum _n |\lambda _n^+|^2 \frac{-\Sigma ^{v}_{n+}}{2\sqrt{M^{*2}_{n+}+\vec{q}^2}}\frac{1}{q_0 - E_n^+ + i\epsilon} \\
& +|\lambda _n^-|^2 \frac{-\Sigma^{v}_{n-}}{2\sqrt{M^{*2}_{n-}+\vec{q}^2}}\frac{1}{q_0 - E_n^- + i\epsilon} + \cdots, 
\end{split}
\label{eq:OPEm3}
\end{equation}
where $|\lambda _n^\pm|^2$ are the residues of the $n$-th states. 
The phenomenological side of the parity projected correlation functions can thus be expressed as follows: 
\begin{equation}
\begin{split}
\Pi ^{\pm}_{m} (q_0, |\vec{q}|) &=  \sum_{n} \frac{|\lambda_{n}^{\pm2}|}{2\sqrt{M^{*2}_{n\pm}+\vec{q}^2}} \frac{(\sqrt{M^{*2}_{n\pm} +\vec{q}^2} + M^{*}_{n\pm})}{q_0 - E^\pm _n + i\epsilon} \\ 
& + \frac{|\lambda_{n}^{\mp2}|}{2\sqrt{M^{*2}_{n\mp}+\vec{q}^2}} \frac{(\sqrt{M^{*2}_{n\mp}+\vec{q}^2}  - M^{*}_{n\mp})}{q_0 - E^\mp _n + i\epsilon} +\cdots, 
\end{split}
\label{eq:mSPFpm}
\end{equation}
We next fit the combinations $\Pi_{m1} (q_0, |\vec{q}|)+\Pi_{m2} (q_0, |\vec{q}|)$, $\Pi_{m1} (q_0, |\vec{q}|) -\Pi_{m2} (q_0, |\vec{q}|)$ 
and $\Pi_{m3} (q_0, |\vec{q}|)$ to the respective OPE functions to extract the effective masses $M^{*2}_{\pm}$ and the vector self-energies 
$\Sigma_{\pm}^{v}$. To be more precise, we substitute the imaginary parts of Eqs.\,(\ref{eq:OPEm1}-\ref{eq:OPEm3}) into Eq.\,(\ref{eq:Gmatter12u}), 
compute the (trivial) $q_0$ integral and fit the result to $G_{m1\mathrm{OPE}}(s, \tau, \theta) + G_{m2\mathrm{OPE}}(s, \tau, \theta)$, 
$G_{m1\mathrm{OPE}}(s, \tau, \theta) - G_{m2\mathrm{OPE}}(s, \tau, \theta)$ and $G_{m3\mathrm{OPE}}(s, \tau, \theta)$. 
To carry out this fit, we keep $E_0^{\pm}$ and $|\lambda_0^{\pm}|^2$ fixed to the values obtained from the MEM analysis of 
$G_{m\mathrm{OPE}}^{\pm}(s, \tau, \theta)$. Specifically, $E_0^{\pm}$ is taken at the energy of the peak maximum and $|\lambda_0^{\pm}|^2$ 
is obtained by integrating the spectral function in the region of the corresponding peak. 
The remaining parameters that need to be fitted are then the factors 
$\frac{E_0^{\pm} +M^*_{0\pm}}{2\sqrt{M^{*2}_{0\pm}+\vec{q}^2}}$ and $-\frac{\Sigma ^{v}_{0+}}{2\sqrt{M^{*2}_{0\pm}+\vec{q}^2}}$, 
from which we can extract the effective masses and vector self-energies. 

In the above fit, one also needs to take the continuum states [not shown in Eqs.\,(\ref{eq:OPEm1}-\ref{eq:OPEm3})] into account. 
For this purpose, we regard the 
continuum obtained from the MEM analysis of $G_{m\mathrm{OPE}}^{+} (s, \tau, \theta)$ and $G_{m\mathrm{OPE}}^{-} (s, \tau, \theta) $ 
as the continuum contributions from $\Pi_{m1} + \Pi_{m2}$ and $\Pi_{m1} - \Pi_{m2}$, respectively. 
Concretely, we assume the $q_0\geq 1050 \mathrm{MeV}$ ($q_0\geq 1750 \mathrm{MeV}$) region to be the continuum of 
$\Pi_{m1} + \Pi_{m2}$ ($\Pi_{m1} - \Pi_{m2}$). 
We have checked that the choice of the lower boundaries has no strong effects on the fitting results. 
The contribution of the continuum state in $\Pi_{m3}$ may furthermore be neglected 
because there are no perturbative contributions to this term in the high energy limit. 

The fit results are given in Fig.\,\ref{fig:selfenergiesrestframe}. 
The left figure shows the behavior for the positive parity state. 
As the density increases, the effective mass decreases, while the vector self-energy increases. 
The values of the effective mass and vector self-energy at normal nuclear matter density are about 130 MeV and 770 MeV, respectively. 
These findings are qualitatively similar to the results of the previous QCD sum rule analyses \cite{Cohen:1994wm,Drukarev:2010xv}, 
while the magnitude of the in-medium modifications are larger than those in Refs.\,\cite{Cohen:1994wm,Drukarev:2010xv}. 
The right figure shows the negative parity effective mass and self-energy. 
The density dependences of both quantities clearly turn out to be much weaker than those of the positive parity state. 

The obtained spectral function, effective mass and vector self-energy for the negative parity state differ 
from the predictions of the chiral doublet models \cite{Detar:1988kn,Jido:1998av}. 
The models predict that 
the mass difference between the nucleon and its negative parity excited state is 
reduced at finite density as it is proportional to the chiral condensate $\langle \overline{q}q \rangle$. 
The models furthermore predict that both the masses monotonically decrease and finally become degenerate in the chirally restored phase. 
Therefore, one expects in this framework that the energy of the negative parity excited state moves towards the positive-parity state 
as the density increases. 
Our study, however shows that the energy of the negative-parity excited state is almost density independent. 
The disagreement between the results of the chiral doublet model and QCD sum rules can be traced back to the 
$\langle q^\dagger q \rangle$ condensate at finite density. 
The cancellation between the changes of $\langle \overline{q} q \rangle$ and $\langle q^\dagger q \rangle$ in medium 
leaves the correlation function (almost) unchanged for the negative-parity sum rule.  
The behavior of the negative-parity nucleon in medium may be experimentally studied from $\eta$-mesic nuclei 
since their structures may be sensitive to the difference between the energies of the nucleon ground state and its first negative parity excitation \cite{Jido:2002yb,Nagahiro:2008rj}. 
It will be interesting to see whether such an experiment can discriminate between the above two pictures and whether it can 
determine which of the two is realized in nature. 

\section{Discussion\label{sec: Discussion}} 
In this section, we discuss the uncertainties of the in-medium condensates and their effects on the sum rule analysis results. 
As we have mentioned in subsection \ref{subsec: construction}, in-medium condensates are evaluated in this work within the linear density approximation 
and their (linear) density dependencies are determined by the values of the quark mass, parton distribution functions etc. 
The values of these quantities have some uncertainties. 
Furthermore, the in-medium values of the higher-order condensates such as the four quark condensates, which are usually evaluated using the factorization hypothesis, 
are poorly known because factorization can only be justified in the large $N_c$ limit. 
We also discuss the spatial momentum dependence of the results and examine the validity of the parity projection for the 
finite momentum case. 

\subsection{Dependence on the in-medium four quark condensates \label{subsec: in-medium four quakr condensate dependence}} 
Four quark condensates can, just like the chiral condensate, 
be related to the spontaneous breaking of chiral symmetry. 
Their contributions to the OPE expression of the correlation function are given in Eqs.\,(\ref{eq:Pi1_for_s}-\ref{eq:Piu_for_s}). 
In the case of the nucleon QCD sum rules in vacuum, the four quark condensates give the 
dominant non-perturbative contribution to the chiral even part $\Pi_1 (q^2)$. 
The in-medium values of the four quark condensates are only poorly constrained because we at present have to rely on 
the factorization hypothesis, according to which the four-quark condensates are given by the square of the chiral condensate. 
This hypothesis may not be justified even in vacuum, while its validity at finite density is even more questionable \cite{Chung:1984gr,Kwon:2008vq,Gubler:2015yna}. 
The density dependence of the four quark condensates and their effects on 
nucleon properties have been studied previously in Refs.\,\cite{Chung:1984gr,Thomas:2007gx,Drukarev:2012av,Jeong:2012pa}, 
but its effect on the lowest negative parity excited state 
is worked out here for the first time. 

In Eqs.\,(\ref{eq:Pi1_for_s}-\ref{eq:Piu_for_s}), three kinds of four quark condensates, namely scalar-scalar $\langle \overline{q}q \rangle ^2$, 
scalar-vector $\langle \overline{q}q \rangle \langle q^{\dagger }q \rangle $ and vector-vector 
$\langle q^{\dagger }q \rangle ^2 $ four quark condensates appear. 
Our integral kernel of Eq.(\ref{eq:phaserokernel2}), in fact, eliminates the contributions of $\langle \overline{q}q \rangle \langle q^{\dagger }q \rangle $ 
and $\langle q^{\dagger }q \rangle ^2 $ at leading order in $\alpha_s$. 
The $\alpha_s$ corrections of these contributions are not considered in this study because 
the $\langle \overline{q}q \rangle \langle q^{\dagger }q \rangle $ and $\langle q^{\dagger }q \rangle ^2 $ condensates 
are not expected to have large contributions up to normal nuclear matter density and thus their $\alpha_s$ corrections presumably are numerically small. 
We therefore study only the effects of the in-medium modification of the scalar-scalar four quark condensates to both the positive and negative parity nucleon states. 
Since only chiral invariant four quark condensates appear in the nucleon QCD sum rule with the Ioffe current [$\beta =-1$ in Eq.(\ref{eq:interpolating field})] 
\cite{Thomas:2007gx}, one could expect that the medium modification of the $\langle \overline{q}q \rangle ^2$ condensate 
may also be small in the vicinity of the Ioffe current (we use $\beta = -0.9$). 
Previous studies actually pointed out that a small density dependence of the four quark condensate
causes realistic results with a slightly decreasing total energy of the positive parity ground 
state, consistent with our knowledge of nuclear phenomenology \cite{Cohen:1994wm}. 
Therefore, to test this possibility, we here assume the $\langle \overline{q}q \rangle ^2$ condensates to be density independent, 
repeat the previous analysis and compare the two results. 
\begin{figure*}[!tbp]
\begin{center}
  \includegraphics[width=10.2cm]{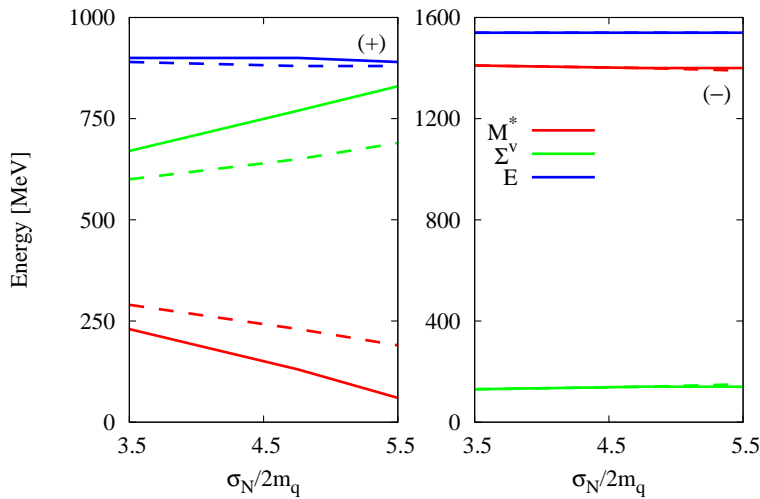}
 \end{center}
 \caption{The $\sigma_{N}$/$2 m_q$ dependence of the effective masses and vector self-energies of 
positive (left) and negative parity (right) states. 
The red and green lines correspond to the effective masses and vector self-energies, respectively. 
The solid lines show the results with full density dependence according to 
the factorization hypothesis for the four quark condensates, while the dashed lines correspond to the 
density independent $\langle \overline{q} q \rangle^2$ case. 
}
 \label{fig:sigma-dependence_rest}
\end{figure*}

This comparison is shown in Fig.\,\ref{fig:selfenergiesrestframe} as the dashed lines. 
In these plots, we see that 
the density dependence of the $\langle \overline{q}q \rangle ^2$ condensate mainly affects the self-energies of the positive parity state. 
The density independent four quark condensate causes 
the effective mass to increase while the vector self-energy decreases. 
On the other hand, the behavior of the negative parity state remains almost completely unchanged. 

\subsection{Dependence on the in-medium chiral condensates \label{subsec: in-medium chiral condensates dependence}}
As we have seen in Fig.\,\ref{fig:OPE density}, the chiral condensate  $\langle \overline{q} q \rangle _m$ term contributes dominantly to 
the phase-rotated QCD sum rule of the nucleon. Its density dependence at the leading order in nucleon density 
is determined by the ratio of $\pi$N sigma term to the light quark mass, $ \xi = \frac{\sigma_N}{2m_q}$ are taken 
as $\sigma_N =45 \, \mathrm{MeV}$, $m_q = 4.725 \, \mathrm{MeV}$ and $\xi \cong 4.76$ in Sec.\ \ref{Sec: result}. 
However, both $\sigma_N$ and $m_q$ have some uncertainties and these precise values are not well determined 
\cite{Shanahan:2012wh,Alarcon:2011zs,Durr:2011mp,Blossier:2010cr,Durr:2010vn,Aoki:2010dy,Hoferichter:2015dsa,Durr:2015dna}. 
Especially for the $\sigma_{N}$ value, a recent dispersion analysis of $\pi N$ scattering data gives a rather large value of 
$\sigma_{\pi N} =  (59 \pm 1.9 \pm 3.0) \mathrm{MeV}$ \cite{Hoferichter:2015dsa}, while another recent lattice calculation 
that uses quark masses at the physical point obtains a much smaller value of $\sigma_{\pi N} = 38(3)(3) \mathrm{MeV}$ \cite{Durr:2015dna}. 

To check the dependence of the self-energies on their values, 
we additionally consider two cases, namely $\xi = 3.5, 5.5$, and show the results in Fig.\,\ref{fig:sigma-dependence_rest}. 
The dependence on the factorization hypothesis for the four quark condensates is also given in this figure. 
One sees that the uncertainty of $\xi$ mainly affects the effective mass and vector self-energy of the positive parity state 
regardless of the density dependence of the four quark condensate. 
The values of the effective mass and vector self-energy at $\xi = 3.5$ and $5.5$ are changed by about 100 MeV and 70 MeV, respectively. 
On the other hand, the other quantities, namely the total energies of both parity states and $M^{*}$ and $\Sigma_v$ of the negative parity state, appear to be 
fairly insensitive to $\xi$. 
The change of $\xi$ also affects the heights of the first peaks in the spectral functions of 
both the positive and negative parity states and thus the values of the respective residues are modified.  

\subsection{Dependence on three-dimensional momentum} 
As a last point, we investigate in this subsection the 
spatial momentum dependencies of the nucleon and its negative parity excited state. 
In the previous sections, we have so far carried out the analyses at rest relative to nuclear matter while 
we shall next study the density dependence of the total energies, effective masses and vector self-energies of both positive and negative parity states 
at the Fermi momentum $|\vec{q}_f|= (\frac{3\pi ^2 \rho _N}{2})^{\frac{1}{3}}$. 
  \begin{figure*}[!tbp]
\begin{center}
  \includegraphics[width=10.2cm]{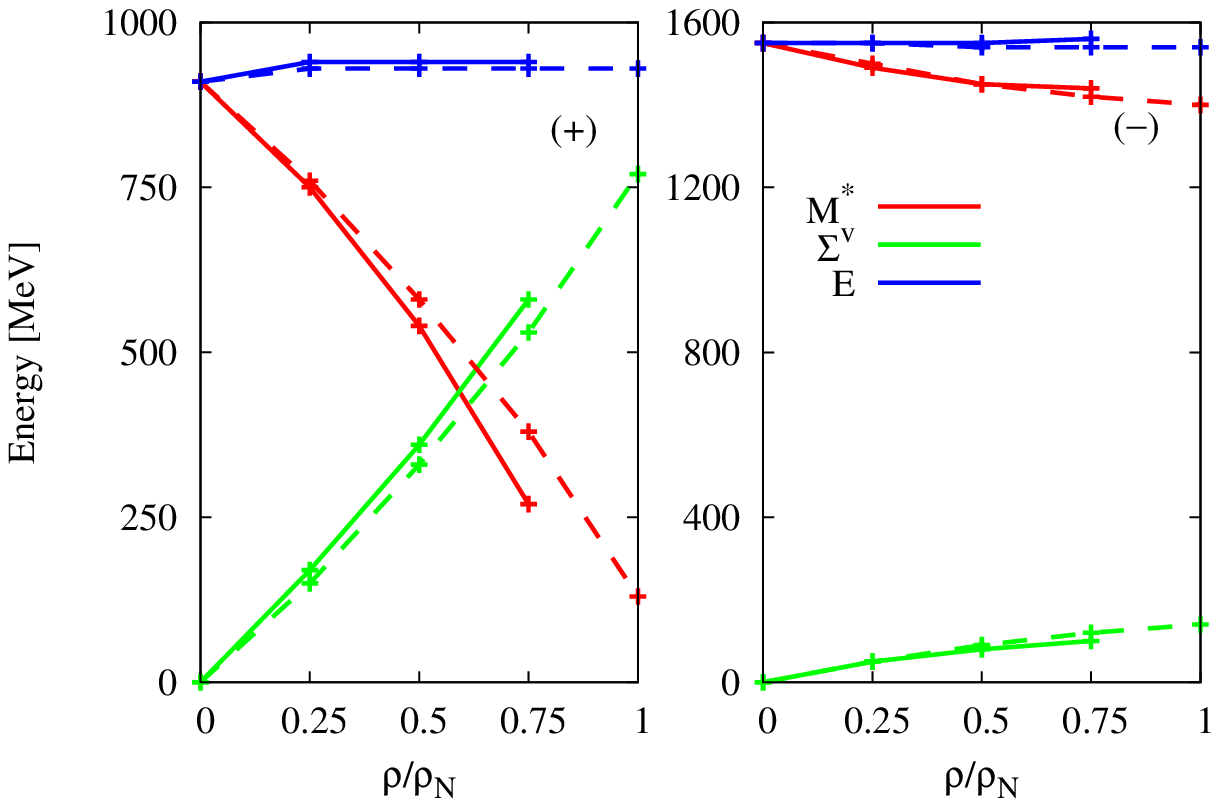}
 \end{center}
 \caption{The density dependence of the effective masses and vector self-energies of 
positive (left) and negative parity (right) states at Fermi momentum. 
The red, green and blue lines correspond to the effective masses, vector self-energies and total energies, respectively. 
For comparison, their values at $|\vec{q}|=0$ are shown as dashed lines. 
}
 \label{fig:selfenergiesfermimomentum}
\end{figure*}
\begin{figure*}[!tbp]
\begin{center}
  \includegraphics[scale =0.6]{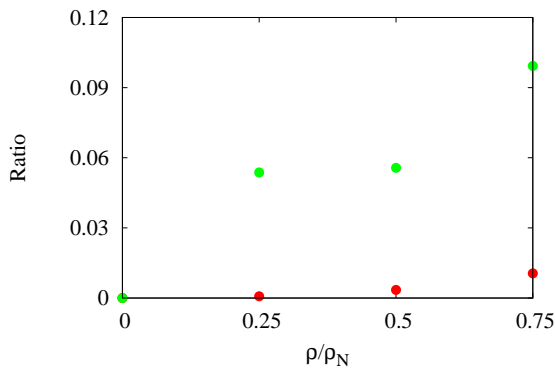}
 \end{center}
 \caption{The red (green) point corresponds to the ratio of the coefficient of the second term to its of first term of 
$\Pi ^{+}_{m} (q_{0}, |\vec{q}|)$ ($\Pi ^{-}_{m} (q_{0}, |\vec{q}|)$) in Eq.(\ref{eq:mSPFpm}). 
}
 \label{fig:factor}
\end{figure*}

The results are shown in Fig.\,\ref{fig:selfenergiesfermimomentum} as solid lines, while 
the dashed lines correspond to the case of the zero spatial momentum. 
The momentum dependencies of the total energies, effective masses and vector self-energies of 
positive and negative parity states turn out to be small. 
The solid curves are not extended to the 
normal nuclear matter density because the MEM analysis of the positive parity states does not work well above $\rho = 0.75 \rho_N$. 
The reason for this failure is 
the rapid decrease of the ground state residue, which is 
faster than for $|\vec{q}| = 0$, making it impossible to extract the positive parity spectral function at $\rho = \rho_N$. 
(The lines of the negative parity state similarly can not be extended to the normal nuclear matter density 
because, when extracting the values of the effective mass and vector self-energies, both of the positive and negative parity spectral function 
are needed.)
The $|\vec{q}|$ dependence of the OPE part originates from the Wilson coefficients which 
depend on the two variables $q^2$ and $q\cdot u$. 
Only terms involving $q \cdot u$ 
lead to $|\vec{q}|$ dependencies of the in-medium spectral functions. 
Such contributions are small up to $0.75 \rho_N$ and 
hence the solid and dashed curves in Fig.\,\ref{fig:selfenergiesfermimomentum} show qualitatively the same behavior. 
Therefore, the weak $|\vec{q}|$ dependencies of the total energies, effective masses and vector self-energies are consistent with the OPE side of the correlation function. 
To investigate the nucleon properties in more detail, higher order contributions to the Wilson coefficients and the condensates would be needed, 
which is a task that goes beyond the scope of this work. 

Finally, we comment on the validity of the parity projection for the non-zero momentum case. 
As can be understood from Eq.\,(\ref{eq:mSPFpm}) the parity projection can only be carried out exactly at $|\vec{q}| =0$, 
meaning that $\Pi^{\pm}_{m}(q_0,|\vec{q}|)$ only receives contributions of states with fixed parity. 
On the other hand, the 
obtained spectral function $\rho^{\pm}_{m\mathrm{Phys.}}(q_0, |\vec{q}|\neq0 )$ may involve some contributions of the opposite parity states 
and thus the peak positions, the effective masses and vector self-energies can in principle be affected by such mixings. 
The contamination can be estimated from the coefficient of the second term in Eq.\,(\ref{eq:mSPFpm}). 
The ratios of the coefficient of the second term to its of first term, namely 
$\frac{|\lambda_{n}^{\mp2}|}{2\sqrt{M^{*2}_{n\mp}+\vec{q}^2}} (\sqrt{M^{*2}_{n\mp}+\vec{q}^2}  - M^{*}_{n\mp})/
\frac{|\lambda_{n}^{\pm2}|}{2\sqrt{M^{*2}_{n\pm}+\vec{q}^2}} (\sqrt{M^{*2}_{n\pm} +\vec{q}^2} + M^{*}_{n\pm}) $, are shown in Fig.\,\ref{fig:factor}. 
In this figure, one observes that for both positive and negative parity states the first term is much larger than the second and thus 
the mixing effect can in practice be ignored for momenta around the Fermi surface. 

\section{Summary and conclusion \label{Sec: summary}}
We have studied the spectral functions of the nucleon and its negative parity excited state in nuclear matter using 
QCD sum rules and the maximum entropy method. 
All known first order $\alpha_s$ corrections to the Wilson coefficients are taken into account and 
the density dependences of the condensates are treated within the linear density approximation. 
With these inputs, we have constructed the parity-projected in-medium nucleon QCD sum rules and have analyzed them with MEM. 
As a result, we have found that the density dependences of the OPE parts are dominated by those of 
the chiral condensate $\langle \overline{q} q \rangle$ and vector quark condensate $\langle q^{\dagger} q \rangle$. 
The difference between the positive and negative parity OPE expressions 
is mainly caused by the chiral condensate term, 
whose sign depends on the parity of the respective nucleon state. 
Therefore, the density dependences of the positive and negative parity OPE parts are rather different. 
As the density increases, the positive parity OPE decreases rapidly because 
the in-medium modifications of $\langle \overline{q}q \rangle$ and $\langle q^{\dagger} q \rangle$ 
are added up to reduce the OPE part. 
On the other hand, the negative parity OPE depends little on the
density due to the cancellation of these modifications. 

We have analyzed these OPE data by MEM and extracted the spectral functions
of positive and negative parity, which in the vacuum exhibit sharp peaks near the
experimental values of the lowest lying states. 
The positions of these peaks turned out to be almost density independent, which means that the total energies of both 
the positive and negative-parity states are not 
modified by nuclear matter effects up to normal nuclear matter density. 
On the other hand, 
as the density increases, the residue of the positive parity nucleon ground state decreases while 
that of the negative parity first excited state remains almost unchanged. 

Assuming mean-field type phenomenological nucleon propagators, we have next investigated the density dependences of the 
effective masses and vector self-energies. 
For positive parity, we have found that 
as the density increases the effective mass decreases while the vector self-energy increases. 
For negative parity, 
the medium modifications of these quantities are very small. 
We have examined potential effects of the uncertainties of the input parameters, namely, the in-medium four quark condensates $\langle \overline{q}q \rangle _m ^2$ 
and the chiral condensate $\langle \overline{q}q \rangle _m $, on the results of the analyses. 
It is found that these uncertainties mainly affect the effective mass and vector self-energy of the positive-parity ground state. 
For larger in-medium modifications of these condensates, 
the effective mass and vector self-energy become more pronounced. 
These results suggest that the effective mass and vector self-energy are strongly correlated to the 
partial restoration of chiral symmetry. 
For negative parity, the in-medium modifications are not much affected by the density dependences of both $\langle \overline{q}q \rangle$ 
and $\langle \overline{q}q \rangle ^2$, which suggests that the results 
qualitatively do not depend on our specific choices for the in-medium condensates. 
We have also investigated the spatial momentum dependence of the nucleon spectra and found that 
the $|\vec{q}|$ dependence of the total energies, the effective masses and vector self-energies of both 
positive and negative parity are small at low density. 
Here, we have restricted ourselves to momenta up to the Fermi momentum at normal nuclear matter density. 
We have also discussed the validity of the parity projection at finite $|\vec{q}|$ and showed that, even though parity projection 
is not exact at finite $|\vec{q}|$, the mixing contributions of 
opposite parity states are sufficiently small up to the Fermi momentum. 

The behaviors of the positive and negative parity states 
can be attributed mainly to the modifications of the $\langle \overline{q}q \rangle$ and $\langle q^{\dagger} q \rangle $ condensates and thus 
our results indicate that both the $\langle \overline{q} q \rangle$ and $\langle q^{\dagger} q \rangle $ condensates are important in 
describing the in-medium properties of the nucleon and its negative parity excited state. 
It is however difficult to provide an intuitive physical interpretation for these findings. 
Our results show that the spectral function, effective mass and vector self-energy of the negative parity state are not modified significantly up to 
normal nuclear matter density. 
These behaviors differ from those obtained from the chiral doublet model, which predicts that 
the effective masses of both the positive and negative parity states decrease monotonically and finally become degenerate in the chirally restored phase. 
As a further point, the decrease of the peak height of the positive-parity spectral function indicates 
that the coupling strength of the nucleon ground state to the interpolating field is reduced rapidly as the density increases. 
These new features of the direct application of QCD are quite interesting, 
while their physical picture is not yet clear and requires further investigation. 

\begin{acknowledgments}
This work is partially supported by KAKENHI under Contract No. 25247036. 
K. O. gratefully acknowledges the support by the Japan Society
for the Promotion of Science for Young Scientists (Contract
No. 25.6520). 
The numerical calculations of this study have been partly performed on the super grid 
computing system TSUBAME at Tokyo Institute of Technology.
\end{acknowledgments}

\end{document}